\documentclass[12pt]{article}
\usepackage[english]{babel}
\usepackage[utf8]{inputenc}
\usepackage[T1]{fontenc}
\usepackage{amsmath}
\usepackage{amsfonts}
\usepackage{url}
\usepackage[dvips]{graphicx}
\usepackage{calc}
\usepackage{subfigure}
\usepackage{multirow}
\def\tildeg{\widetilde{g}}
\def\mL{\mathcal{L}}
\def\mH{\mathcal{H}}
\def\tf{\tilde{f}}
\def\tN{\tilde{N}}
\def\tp{\tilde{p}}
\def\tphi{\tilde{\phi}}
\newcommand{\half}[1]{\frac{#1}{2}}
\newcommand{\halb}{\half{1}}

\newcommand{\inv}[1]{\frac{1}{#1}}
\newcommand{\dd}{\:\text{d}}

\newcommand{\kzav}[1]{\left(#1\right)}
\newcommand{\hzav}[1]{\left[#1\right]}

\newcommand{\pd}[2]{\frac{\partial #1}{\partial #2}}

\newcommand{\rf}[1]{(\ref{#1})}

\begin{document}
\begin{center}
{\Large{ \bf Einstein and Jordan-Frame  Covariant Hamiltonians for $F(R)$ Gravity and Their Canonical Relationships}}

	\vspace{1em}  J. Kluso\v{n}\textsuperscript{\textdagger} and B. Matou\v{s}\textsuperscript{\textdagger}\textsuperscript{\textdaggerdbl}
			\footnote{Email addresses:
			J. Kluso\v{n}:\ klu@physics.muni.cz, B. Matou\v{s}:\  bmatous@mail.muni.cz}\\
			\vspace{1em} \textsuperscript{\textdagger} \textit{Department of Theoretical Physics and
				Astrophysics, Faculty
				of Science,\\
				Masaryk University, Kotl\'a\v{r}sk\'a 2, 611 37, Brno, Czech Republic}\\
			\vspace{1em} \textsuperscript{\textdaggerdbl} \textit{North-Bohemian Observatory and Planetarium in Teplice, \\
Kopern\'{i}kova 3062, 415 01, Teplice,
Czech Republic}\\

\end{center}

\abstract{This paper is devoted to the analysis of the covariant 
canonical formalism of $F(R)$ gravity in Einstein frame. We also find  canonical transformation between covariant canonical formulation of $F(R)$ gravity in   Jordan frame and Einstein frames and we also determine corresponding generating function.}

\section{Introduction and Summary}
Fundamental objects in relativistic theory are fields whose dynamics is derived from the action principle. The action is manifestly covariant object that is defined as the 
space-time integral of Lagrangian density. This manifest covariance is lost in the Hamiltonian formalism since its crucial point is an existence of  one preferred coordinate which is the time coordinate.
There is an alternative Hamiltonian formalism, known as 
 covariant Hamiltonian theory or the Weyl-De Donder theory \cite{DeDonder, Weyl} that maintains manifest covariance. In the covariant Hamiltonian theory, the momenta are defined as derivatives with respect to all partial derivatives of coordinates so that there is no preferred direction and hence manifest covariance is preserved. This is very attractive idea that could be especially useful in manifest covariant theories as for example theory of gravity.  In fact, the first covariant Hamiltonian theory of  General Relativity was published by Ho\v{r}ava long time ago \cite{Horava:1990ba}. Recently, this work was further discussed examining its thermodynamic consequences in \cite{Parattu:2013gwa}.

The obvious next step is to apply this approach to some generalized theories of gravity. The simplest one is the $F(R)$ gravity  where the scalar curvature in the Lagrangian  is replaced by a general function $F(R)$ of scalar curvature
\footnote{For review of this theory see \cite{DeFelice:2010aj, Nojiri:2017ncd, Nojiri:2010wj}.}. The covariant Hamiltonian for $F(R)$ gravity in Jordan frame was found in \cite{Kluson:2020tzn}.

It is well known that $F(R)$ gravity can also be formulated in Einstein frame,
for detailed discussion, see for example \cite{DeFelice:2010aj}
and also \cite{Chakraborty:2016ydo,Chakraborty:2016gpg} and for the discussion of equivalence of these two frames, see also
\cite{Capozziello:2006dj,Bahamonde:2017kbs,Bahamonde:2016wmz,Briscese:2006xu}. The transformations between these two frames is based on the Weyl transformation of metric and corresponding Riemann and Ricci tensor. In more details, with appropriate chosen 
Weyl factor we can  arrive to Einstein frame formulation of $F(R)$ gravity. $F(R)$ Lagrangian in Einstein frame is very similar to General Relativity Lagrangian minimally coupled to a scalar field. In 3+1 formalism, Hamiltonians for both frames were formulated and there was found that they are related by canonical transformation \cite{Deruelle:2009pu}. This leads to a question whether this is also true in the covariant Hamiltonian formalism. This question is answered with the presented paper.
In more details, we firstly derive the covariant Hamiltonian for $F(R)$ gravity in Einstein frame. We find that covariant Hamiltonian for $F(R)$ gravity in Einstein frame has the same form as was found in \cite{Horava:1990ba} together with new additional scalar field contribution. On the other hand the covariant Hamiltonian formulation of $F(R)$ gravity in Jordan frame was performed in \cite{Kluson:2020tzn} and our goal is to show that these two formulations are related by canonical transformations. To do this we should firstly examine how canonical transformations are defined in covariant Hamiltonian formalism, following 
\cite{Struckmeier:2008zz,Struckmeier:2012de}. Then we study 
 relationship between covariant Hamiltonians for $F(R)$ gravity in  Jordan  and Einstein frames and we find  generating function
 resulting in proof of canonical transformation between those two Hamiltonians. This is really new and non-trivial result that shows that these two Hamiltonians are related by canonical transformations in the similar way how  two Hamiltonians are related in $3+1$  canonical formalism \cite{Deruelle:2009pu}. On the other hand there is crucial difference between these two canonical transformations which is in the preservation of the Poisson brackets. In fact, it is not completely clear how to define Poisson brackets in covariant canonical formalism due to the fact that the conjugate momenta have additional vector index. Then it is natural to define Poisson bracket as in \cite{Struckmeier:2008zz} where the Poisson bracket is defined in the same way as canonical Poisson bracket so that it is now labeled by vector index. It was shown in \cite{Struckmeier:2008zz} that such Poisson brackets are not generally preserved under canonical transformations and we show that exactly this situation occurs in case of canonical transitions between Einstein and Jordan frame covariant Hamiltonians for $F(R)$-gravity. On the other hand we show that the form of Lagrangian brackets is preserved under canonical transformation. 
 We mean that this is very interesting result that demonstrates nice application of the covariant canonical formalism for the study of $F(R)$-gravity.

The organization of this paper is as follows. 
In the next section (\ref{second}) we  formulate  covariant Hamiltonian for $F(R)$ gravity in Einstein frame. The  section (\ref{third}) is devoted to the canonical transformation between these two frames. At first the transformation's existence is proven using fundamental Lagrange brackets, then the explicit form of the generating function of this transformation is found, and at the end the Poisson brackets are also noted, they role is however a minor one since in the covariant Hamiltonian theory they do not serve as canonical invariant. The fourth section (\ref{fourth}) deals with the surface term of the Lagrangian
\footnote{
This paper uses the East Coast convention with metric signature ($-,+,+,+$) and
Latin indices running over $0...3$ interval while the Greek ones over $1...3$. The fundamental
constants $c, G, \hbar, k_B$ are treated as equal to one. }.

\section{$F(R)$-gravity in Einstein Frame and Its Covariant Hamiltonian}\label{second}
We begin this section with the introduction of the
Lagrangian for $F(R)$ gravity. 
$F(R)$ gravity is the simplest generalization of the Einstein-Hilbert
action when we replace the linear dependence of the Lagrangian 
density on the scalar curvature~$R$ by more general function 
$F(R)$ \footnote{For review and extended list of references, see for example \cite{DeFelice:2010aj}.}. Explicitly, the Lagrangian density of $F(R)$ theory of gravity has the form 
\begin{equation}\label{lagFR}
    \mathcal{L} = \frac{\sqrt{-g}}{16\pi}F(R) \:.
\end{equation}
The presence of the function $F(R)$ implies that it is not straightforward procedure to find corresponding Hamiltonian. In order to overcome this issue it is 
 convenient to introduce two scalar fields $A$ and $B$ and replace the Lagrangian density (\ref{lagFR}) by the following one 
\begin{equation}\label{FRextend}
\mathcal{L} = \frac{\sqrt{-g}}{16\pi} \hzav{F(B) + A\kzav{R-B}}\ . 
\end{equation}
In fact, equations of motion for $A$ and $B$ that follow from
(\ref{FRextend}) have the form 
\begin{equation}\label{eqFRexed}
R-B=0  \ , \quad F'(B)-A=0 \ , \quad  F'(B)\equiv \frac{dF}{dB} \ . 
\end{equation}
Then inserting the first equation in (\ref{eqFRexed}) 
into (\ref{FRextend}), we easily see the Lagrangian density (\ref{FRextend}) reduces into (\ref{lagFR}) that shows equivalence of these two actions. For our purposes, it is useful to use the second equation in (\ref{eqFRexed}) to solve $A$ as function of $B$ and hence the Lagrangian density in Jordan frame has the form
\begin{equation}
 \mathcal{L}^J = \frac{\sqrt{-g}}{16\pi} \hzav{F(B) + F'(B)\kzav{R-B}}\:.
\end{equation}
As it is well known, we can formulate $F(R)$-gravity in the Einstein frame too, see for example \cite{DeFelice:2010aj}. Note that the Einstein frame is defined by requirement that the Lagrangian density is linear in the scalar curvature $R$. In the rest of this paper, the coordinates with a tilde are coordinates in Einstein frame while those without punctuation belong to Jordan frame. In order to find  the transformation from Jordan frame to Einstein one we should perform  Weyl transformation of the metric that is defined as
\begin{equation}\label{trang}
 \widetilde{g}_{ij} = F' g_{ij}\ . 
\end{equation}
As the next step we  introduce connection with tilde 
\begin{equation}
\widetilde{\Gamma}^i_{jk}=\frac{1}{2}\tildeg^{il}
(\partial_j \tildeg_{lk}+\partial_k \tildeg_{lj}-\partial_l \tildeg_{jk})  \ . 
\end{equation}
Then using (\ref{trang}) we find how it is related to $\Gamma^i_{ij}$ and to $F(B)$
\begin{equation}
\label{tGamma}
 \widetilde{\Gamma}^i_{jk} = \Gamma^i_{jk} + \halb \delta^i_j \partial_k (\ln F') + \halb \delta^i_k \partial_j (\ln F') - \halb g_{jk} g^{km}\partial_m (\ln F')\:.
\end{equation}
With the help of this relation between connections, we can easily find relation between corresponding scalar curvatures and we get
\begin{equation}
 \widetilde{R} = \widetilde{g}^{ab} \widetilde{R}_{ab} = \inv{F'}R - \frac{3}{F'} \Box \ln F' - \frac{3}{2F'} \partial_i \ln F' g^{ij} \partial_j \ln F'
  \:.
\end{equation}
For writing the Lagrangian in Einstein frame, we need to express $R$ in terms of $\widetilde{R}$, $\widetilde{g}_{ij}$ and corresponding derivatives. To do this, we use the following rules 
\begin{equation}
\Box \ln F' = F' \widetilde{\Box} \ln F' - F' \kzav{\widetilde{\partial}\ln F'}^2 \:,
\quad
\kzav{\partial \ln F'}^2 = F'\kzav{\widetilde{\partial}\ln F'}^2 \ .
\end{equation} 
Then we can easily express $R$ as 
\begin{equation}
 R = F' \hzav{ \widetilde{R} + 3 \widetilde{\Box} \ln F' - \frac{3}{2} \kzav{\widetilde{\partial} \ln F'}^2}
  \:.
\end{equation}
Using this expression together with the fact that $\sqrt{-\widetilde{g}} = F'^2 \sqrt{-g}$ we obtain that the Lagrangian density for $F(R)$ gravity takes the form
 
\begin{equation}
 \mathcal{L} = \frac{\sqrt{-\widetilde{g}}}{16\pi} 
 \hzav{
 \frac{F - F'B}{F'^2} + 
 \widetilde{R} + 
 3 \widetilde{\Box} \ln F' 
 -\frac{3}{2} \kzav{\widetilde{\partial} \ln F'}^2
 }
 \:,
\end{equation}
or equivalently
\begin{equation}
 \mathcal{L} = \frac{\sqrt{-\widetilde{g}}}{16\pi} 
 \hzav{
 \frac{F - F'B}{F'^2} + 
 \widetilde{R} 
 -\frac{3}{2} \kzav{\widetilde{\partial} \ln F'}^2
 } +  \frac{3}{16\pi}\partial_i \kzav{\sqrt{-\widetilde{g}} \widetilde{\partial}^i \ln F'} 
 \:.
\end{equation}
In order to get canonical form of the Lagrangian density, 
we introduce a new scalar variable $\widetilde{\phi} = \sqrt{\frac{3}{2}} \ln F'$ and a potential $V(B) = \frac{F'B - F}{F'^2}$
so that the Lagrangian density takes Einstein-Hilbert form
\begin{equation}
 \mathcal{L}^E = \frac{\sqrt{-\widetilde{g}}}{16\pi} 
 \hzav{
 \widetilde{R} 
 -\tildeg^{ij}\partial_i\widetilde{\phi}\partial_j
 \widetilde{\phi}
 - V(\widetilde{\phi})
 } +  \frac{\sqrt{6}}{16\pi}\partial_i \kzav{\sqrt{-\widetilde{g}}\tildeg^{ij} \partial_j \widetilde{\phi}} 
 \:.
\end{equation}
This is the final form of Lagrangian density for $F(R)$ gravity in Einstein frame. To proceed to the covariant canonical formalism, it is necessary to separate  Lagrangian  into two parts
\cite{Horava:1990ba}: the bulk term that contains only the first derivatives and the surface term which can be expressed as a total  derivative. To construct Hamiltonian, solely the bulk term is needed, the surface Hamiltonian will be discussed in section (\ref{fourth}). Explicitly, we get
\begin{eqnarray}\label{mLbulk}
&&\mL^E=\mL^E_{bulk}+\mL^E_{sur} \ , \nonumber \\
 &&\mathcal{L}_{bulk} = \frac{\sqrt{-\widetilde{g}}}{16\pi} 
 \hzav{
 \widetilde{g}^{ab}\widetilde{\Gamma}^c_{bd}\widetilde{\Gamma}^d_{ac}
 -  \widetilde{g}^{ac}\widetilde{\Gamma}^b_{bd}\widetilde{\Gamma}^d_{ac}
 - \tildeg^{ij}\partial_i \widetilde{\phi}
 \partial_j \widetilde{\phi}
 - V(\widetilde{\phi})
 } \ , \nonumber \\
&&\mathcal{L}_{sur} =  
\partial_i \hzav{\frac{\sqrt{-\widetilde{g}}}{16\pi}
\kzav{\widetilde{g}^{ab}\widetilde{\Gamma}^i_{ab} - \widetilde{g}^{ai}\widetilde{\Gamma}^b_{ab} 
+  \sqrt{6} \tildeg^{ij}\partial_j \widetilde{\phi} }
} \ . \nonumber \\ \label{Lagrangians}
\end{eqnarray}
We  see that the bulk part reminds the bulk Lagrangian of General Relativity \cite{Parattu:2013gwa} together with added $\tphi$-related terms. 

First step to transform the Lagrangian into covariant  Hamiltonian is to find corresponding momenta.
From (\ref{mLbulk}) we  find that the  momentum conjugated to $\widetilde{g}$ is equal to
\begin{equation}\label{Mabc}
\widetilde{M}^{prs} = \frac{\partial \mathcal{L}_{bulk}}{\partial \partial_p \widetilde{g}_{rs}} = 
\frac{\sqrt{-\widetilde{g}}}{16\pi} \hzav{
\widetilde{\Gamma}^p_{ab} 
\kzav{
\widetilde{g}^{ar} \widetilde{g}^{bs}
- \halb \widetilde{g}^{ab} \widetilde{g}^{rs}
}
+ \halb \widetilde{\Gamma}^a_{ak} 
\kzav{
\widetilde{g}^{rs} \widetilde{g}^{kp}
- \widetilde{g}^{ps} \widetilde{g}^{kr}
- \widetilde{g}^{pr} \widetilde{g}^{ks}
}
}
\:.
\end{equation}
However it was shown in 
\cite{Horava:1990ba,Parattu:2013gwa,Kluson:2020tzn}
that the covariant  canonical formalism of general relativity is 
better formulated when we introduce  coordinate $\widetilde{f^{ab}}$
that is related to $\widetilde{g}$ by following relation
\begin{equation}\label{deffab}
\widetilde{f}^{ab} = \sqrt{-\widetilde{g}} \widetilde{g}^{ab}
\ ,
\end{equation}
where, following  \cite{Kluson:2020tzn}, we define $\tf_{ab}$ as  inverse to $\tf^{ab}$
\footnote{This definition is different from the one used  in \cite{Parattu:2013gwa}, where the new coordinate was defined so $f^{ab} f_{cb} = -f\delta^a_c$.}
\begin{equation}
\tf_{ab}\tf^{bc}=\delta_a^c \ . 
\end{equation}
Note that from (\ref{deffab}) we also get useful result
\begin{equation}
\tf \equiv \det \tf^{ab}=\widetilde{g} \ . 
\end{equation}
The momentum $\widetilde{N}^c_{ab}$ conjugated to $\tf^{ab}$ can be obtained directly when $\tf^{ab}$ is substituted into Lagrangian or by using formulae \cite{Kluson:2020tzn}
\begin{equation}
\widetilde{N}^c_{ab} = -\widetilde{M}^{cmn} \inv{\sqrt{-\widetilde{f}}} \widetilde{B}_{mn\:ab}
\:,
\qquad
\widetilde{B}_{mn\:ab} = \halb \kzav{
\widetilde{g}_{ma} \widetilde{g}_{bn}
+ \widetilde{g}_{mb} \widetilde{g}_{an}
- \widetilde{g}_{mn} \widetilde{g}_{ab}
}
\:.
\end{equation}
Using explicit form for $\widetilde{M}^{abc}$ given in (\ref{Mabc}) we obtain the well known result
\begin{equation}
\label{Ncab}
\widetilde{N}^c_{ab} = \inv{16\pi} \hzav{
-\widetilde{\Gamma}^c_{ab} 
+ \halb \kzav{
\widetilde{\Gamma}^k_{ak}  \delta^c_b
+ \widetilde{\Gamma}^k_{kb}  \delta^c_a
}
}
\ .
\end{equation}
From (\ref{mLbulk}) we also find  momentum conjugate to $\widetilde{\phi}$ 
\begin{equation}
\widetilde{p}^{a} = \frac{\partial \mathcal{L}_{bulk}}{\partial \partial_a \widetilde{\phi}} = 
 -\inv{8\pi} \widetilde{f}^{ab} \partial_b \widetilde{\phi}
\:.
\end{equation}
The Hamiltonian is then computed using Legendre transformation as
\begin{eqnarray}
&&\mathcal{H}^E = \partial_c \widetilde{f}^{ab} \widetilde{N}^c_{ab} + \partial_a \widetilde{\phi} \widetilde{p}^a - \mathcal{L}_{bulk}
=\nonumber \\
&&= \frac{1}{16\pi}
\hzav{
\widetilde{f}^{ab}
\kzav{
\widetilde{\Gamma}^c_{db}\widetilde{\Gamma}^d_{ca}
- \widetilde{\Gamma}^c_{ab}\widetilde{\Gamma}^d_{dc}
}
+ \sqrt{-\widetilde{f}} V
-\widetilde{f}^{ab}
\partial_a \widetilde{\phi}
\partial_b \widetilde{\phi}
}
\ . \nonumber \\
\end{eqnarray}
Finally we should express this Hamiltonian in terms of canonical variables which can be done using the relations
\begin{equation}
\label{inverseMomenta}
\widetilde{\Gamma}^i_{jk} = 16\pi 
\hzav{
-\widetilde{N}^i_{jk}
+\inv{3}\kzav{
\widetilde{N}^u_{ju} \delta^i_k
+\widetilde{N}^u_{ku} \delta^i_j
}
}
 \:,
 \quad
 \partial_a \widetilde{\phi} = -8\pi \widetilde{f}_{ab} \widetilde{p}^b
 \:.
\end{equation}
Then the final form of covariant Hamiltonian for Einstein frame of $F(R)$ gravity, is expressed as
\begin{equation}
\label{EFHam}
\mathcal{H}^E =
16\pi\widetilde{f}^{ab} \kzav{
\widetilde{N}^c_{bd} \widetilde{N}^d_{ac}
-\inv{3} \widetilde{N}^c_{ac} \widetilde{N}^d_{bd}
}
+ \frac{\sqrt{-\widetilde{f}}}{16\pi} V
- 4\pi \widetilde{f}_{ab} \widetilde{p}^a \widetilde{p}^b
\:.
\end{equation}

\section{Relationship with Jordan Frame Hamiltonian}\label{third}
Now we proceed to the main part of this paper which is the  relationship between covariant Hamiltonians in Einstein and Jordan  frames respectively. 
Our work is motivated  by an interesting paper 
\cite{Deruelle:2009pu} where the Hamiltonian for $F(R)$ gravity in $3+1$ formalism was analysed and it was shown there that they are related by canonical transformations. 
Then it is very interesting question whether such canonical transformation exists in the case of the covariant canonical formalism too. 

\subsection{Lagrange Brackets}
If there is a canonical transformation between the two frames, the fundamental Lagrange brackets shall be preserved \cite{Struckmeier:2008zz}. Contrary to the conventional Hamiltonian theory, it is not good to use Poisson brackets for this purpose, because fundamental Poisson brackets are preserved only \cite{Struckmeier:2008zz} when the transformed momenta do not depend on original coordinates or on original momenta. This condition is quite strong and, as it will be presented in this section, does not hold for our system. The Poisson brackets will be noted shortly in section \ref{PoissonBrackets}. So the calculation of Lagrange brackets of Jordan frame variables in Einstein frame can reveal us the existence of a canonical transformation. In order to calculate the Lagrange brackets we need relations between Jordan frame variables and Einstein frame ones. We are already able to express the transformation rules for coordinates
\begin{equation}
\label{coordTransformation}
\widetilde{f}^{ab} = F' f^{ab}
\:, \qquad
\widetilde{\phi} = \sqrt{\frac{3}{2}} \ln F'
\:.
\end{equation}
In order to find transformation of momenta $\tN^c_{ab}$, we need to take \rf{Ncab} and \rf{tGamma} and compare it with the the Jordan frame momenta \cite{Kluson:2020tzn}
\begin{equation}
N^c_{ab} = \frac{F'}{16\pi}\hzav{
-\Gamma^c_{ab} 
+ \halb \kzav{\Gamma^k_{ak}\delta^c_b + \Gamma^k_{bk} \delta^c_a} 
+ \halb \kzav{
\delta^c_b \partial_a \ln F' 
+ \delta^c_a \partial_b \ln F' 
+ f^{gc} f_{ab}\partial_g \ln F' 
}
} \:,
\end{equation}
resulting in transformation relation
\begin{equation}
\label{tNinJordan}
 \widetilde{N}^c_{ab} = \inv{F'} N^c_{ab}
 \:,
 \end{equation}
that also implies following important relation
\begin{equation}
\label{fN=tftN}
    \tf^{ab}\tN^c_{ab} = f^{ab}N^c_{ab}\:.
\end{equation}
From the relation \rf{tNinJordan}, we see that the transformed momentum depends on original momentum as well as original coordinate, so the condition for canonical invariance of Poisson brackets is not met. As the next step, we proceed to the transformation of momentum $\tp^a$. First of all, we use the second relation in \rf{inverseMomenta} where we insert \rf{coordTransformation} so that we find  relation between $\partial_a B$ and $\tp^a$. Further, we use the relation between $\partial_a B$ and $p^a$ that was derived in \cite{Kluson:2020tzn} 
\begin{equation}
    \partial_g B = \frac{16\pi}{3F''} f_{gc} \kzav{f^{ik} N^c_{ik} - \frac{F'}{F''} p^c}
    \ . 
\end{equation}
If we combine these relations together we find desired relation
 \begin{equation}
\label{tpinJordan}
 \widetilde{p}^a =\sqrt{\frac{2}{3}}\kzav{
   \frac{F'}{F''}p^a - f^{bc} N^a_{bc}
 }
 \:.
\end{equation}
Having found the transformation relations, the brackets follow simply as
\begin{align}
    &\left\{f^{ab},\:f^{cd}\right\}^j =
    \pd{\widetilde{f}^{ik}}{f^{ab}}\pd{\widetilde{N}^j_{ik}}{f^{cd}}
    -\pd{\widetilde{N}^j_{ik}}{f^{ab}}\pd{\widetilde{f}^{ik}}{f^{cd}}
    +\pd{\widetilde{\phi}}{f^{ab}}\pd{\widetilde{p^j}}{f^{cd}}
    -\pd{\widetilde{p^j}}{f^{ab}}\pd{\widetilde{\phi}}{f^{cd}}
    = 0 \:, \nonumber \\
    &\left\{f^{ab},\:B\right\}^j =
    \pd{\widetilde{f}^{ik}}{f^{ab}}\pd{\widetilde{N}^j_{ik}}{B}
    -\pd{\widetilde{N}^j_{ik}}{f^{ab}}\pd{\widetilde{f}^{ik}}{B}
    +\pd{\widetilde{\phi}}{f^{ab}}\pd{\widetilde{p^j}}{B}
    -\pd{\widetilde{p^j}}{f^{ab}}\pd{\widetilde{\phi}}{B}
    = 0 \:,\nonumber \\
    &\left\{B,\:B\right\}^j = 0 \text{ by definition}
\:, \nonumber \\
    &\left\{f^{ab},\:N^{c}_{de}\right\}^j =
    \pd{\widetilde{f}^{ik}}{f^{ab}}\pd{\widetilde{N}^j_{ik}}{N^{c}_{de}}
    -\pd{\widetilde{N}^j_{ik}}{f^{ab}}\pd{\widetilde{f}^{ik}}{N^{c}_{de}}
    +\pd{\widetilde{\phi}}{f^{ab}}\pd{\widetilde{p^j}}{N^{c}_{de}}
    -\pd{\widetilde{p^j}}{f^{ab}}\pd{\widetilde{\phi}}{N^{c}_{de}}
    =  \delta^j_c \delta^{de}_{ab} \:, \nonumber \\
    &\left\{f^{ab},\:p^{c}\right\}^j =
    \pd{\widetilde{f}^{ik}}{f^{ab}}\pd{\widetilde{N}^j_{ik}}{p^{c}}
    -\pd{\widetilde{N}^j_{ik}}{f^{ab}}\pd{\widetilde{f}^{ik}}{p^{c}}
    +\pd{\widetilde{\phi}}{f^{ab}}\pd{\widetilde{p^j}}{p^{c}}
    -\pd{\widetilde{p^j}}{f^{ab}}\pd{\widetilde{\phi}}{p^{c}}
    = 0 \:, \nonumber \\
    &\left\{B,\:N^{c}_{de}\right\}^j =
    \pd{\widetilde{f}^{ik}}{B}\pd{\widetilde{N}^j_{ik}}{N^{c}_{de}}
    -\pd{\widetilde{N}^j_{ik}}{B}\pd{\widetilde{f}^{ik}}{N^{c}_{de}}
    +\pd{\widetilde{\phi}}{B}\pd{\widetilde{p^j}}{N^{c}_{de}}
    -\pd{\widetilde{p^j}}{B}\pd{\widetilde{\phi}}{N^{c}_{de}}
    = 0 \:, \nonumber \\
    &\left\{B,\:p^{c}\right\}^j =
    \pd{\widetilde{f}^{ik}}{B}\pd{\widetilde{N}^j_{ik}}{p^{c}}
    -\pd{\widetilde{N}^j_{ik}}{B}\pd{\widetilde{f}^{ik}}{p^{c}}
    +\pd{\widetilde{\phi}}{B}\pd{\widetilde{p^j}}{p^{c}}
    -\pd{\widetilde{p^j}}{B}\pd{\widetilde{\phi}}{p^{c}}
    = \delta^j_c \:, \nonumber \\
    &\left\{N^c_{ab},\:N^{d}_{ef}\right\}^j =
    \pd{\widetilde{f}^{ik}}{N^c_{ab}}\pd{\widetilde{N}^j_{ik}}{N^{d}_{ef}}
    -\pd{\widetilde{N}^j_{ik}}{N^c_{ab}}\pd{\widetilde{f}^{ik}}{N^{d}_{ef}}
    +\pd{\widetilde{\phi}}{N^c_{ab}}\pd{\widetilde{p^j}}{N^{d}_{ef}}
    -\pd{\widetilde{p^j}}{N^c_{ab}}\pd{\widetilde{\phi}}{N^{d}_{ef}}
    = 0 \:, \nonumber \\
    &\left\{N^c_{ab},\:p^{d}\right\}^j =
    \pd{\widetilde{f}^{ik}}{N^c_{ab}}\pd{\widetilde{N}^j_{ik}}{p^{d}}
    -\pd{\widetilde{N}^j_{ik}}{N^c_{ab}}\pd{\widetilde{f}^{ik}}{p^{d}}
    +\pd{\widetilde{\phi}}{N^c_{ab}}\pd{\widetilde{p^j}}{p^{d}}
    -\pd{\widetilde{p^j}}{N^c_{ab}}\pd{\widetilde{\phi}}{p^{d}}
    = 0 \:, \nonumber \\
    &\left\{p^{a},\:p^{b}\right\}^j =
    \pd{\widetilde{f}^{ik}}{p^{a}}\pd{\widetilde{N}^j_{ik}}{p^{b}}
    -\pd{\widetilde{N}^j_{ik}}{p^{a}}\pd{\widetilde{f}^{ik}}{p^{b}}
    +\pd{\widetilde{\phi}}{p^{a}}\pd{\widetilde{p^j}}{p^{b}}
    -\pd{\widetilde{p^j}}{p^{a}}\pd{\widetilde{\phi}}{p^{b}}
    = 0 \:, \nonumber \\
\end{align}
where we have used abbreviation $\delta^{ab}_{cd} = \halb \kzav{\delta^a_c \delta^b_d + \delta^b_c \delta^a_d}$.
From the equations above, it is easily visible that the fundamental Lagrange brackets are preserved, so there is a canonical transformation between Jordan and Einstein frames.

\subsection{Generating Function of Canonical Transformation}
Having proven existence of canonical transformation, its generating function is looked for.
But first, let us review basic facts about canonical transformations in covariant formalism, following \cite{Struckmeier:2008zz, Struckmeier:2012de}. 

Let us consider  covariant formulation of $F(R)$ gravity in Jordan frame with  the canonical variables $f^{ab},N_{ab}^c,B,p^c_B$. On the other hand in case of covariant formulation of Einstein-frame $F(R)$-gravity the canonical variables are 
$\tf^{ab},\tN_{ab}^c,\tphi,\tp^a$. We demand that they give the same description of the physical systems so that we have a requirement
\begin{eqnarray}
\delta \int d^4x
(N^c_{ab}\partial_c f^{ab}+p^c\partial_c B-\mH^J)=
\delta\int d^4x
(\tN^c_{ab}\partial_c \tf^{ab}+\tp^c\partial_c \tphi-\mH^E) \ . 
\end{eqnarray}
This result implies that the integrals can differ only by divergence of a vector function whose variation vanishes on the boundary $\partial R$ of the integration region $R$ 
\begin{equation}
\label{CanonicalSurfaceCondition}
    \delta \int_R d^4x \partial_a G_1^a(f,B,\tf,\tphi)=
    \delta \oint_{\partial R}G_1^a(f,B,\tf,\tphi) dS_a=0
    \: .
\end{equation}
Using $G_1^a$, we can write 
\begin{equation}
\label{eqLagrangians}
N^c_{ab}\partial_c f^{ab}+p^c\partial_c B-\mH^J=
\tN^c_{ab}\partial_c \tf^{ab}+\tp^c\partial_c \tphi-\mH^E+
\partial_a G_1^a(f,B,\tf,\tphi,x) \:.
\end{equation}
The divergence of $G_1^a$ can be expressed using coordinates as 
\begin{equation}
    \partial_a G^a_1=
    \frac{\partial G_1^a}{\partial f^{bc}}\partial_a f^{bc}+
    \frac{\partial G_1^a}{\partial \tf^{bc}}\partial_a \tf^{bc}+
    \frac{\partial G_1^a}{\partial B}\partial_a B+
    \frac{\partial G_1^a}{\partial \tphi}\partial_a \tphi+
\frac{\partial G_1^a}{\partial x^a}|_{expl}    \:.
\end{equation}
Putting this into \rf{eqLagrangians} and comparing terms proportional to $\partial_c f^{ab}$ and 
$\partial_c \tf^{ab}$ and $\partial_a B$ and $\partial_a \tphi$ we obtain 
\begin{align}
N^c_{ab} &=\frac{\partial G_1^c}{\partial f^{ab}} \ , 
\quad \quad
p^c =\frac{\partial G_1^c}{\partial B} \ , 
\nonumber \\
\tN_{ab}^c &=-\frac{\partial G_1^c}{\partial \tf^{ab}} \ , 
\quad \:
\tp^c =-\frac{\partial G_1^a}{\partial \tphi} \ , 
\nonumber \\
\mH^J&=\mH^E-\frac{\partial G_1^a}{\partial x^a}|_{expl} 
\:. 
\end{align}
On the other hand we can consider 
different type of generating function 
\begin{equation}
    \label{G2Def}
    G_1^a = G_2^a(f,B,\tN,\tp)-\tf^{ik}\tN_{ik}^a-
\tphi \tp^a \:,
\end{equation}
so that its total divergence is equal to
\begin{eqnarray}
&&    \partial_c (G^c_2
-\tf^{ab}\tN_{ab}^c-
\tphi \tp^c)=\nonumber \\
&&    \frac{\partial G^c_2}{\partial f^{ab}}
    \partial_c f^{ab}+
  \frac{\partial G^c_2}{\partial B}\partial_c B+
    \frac{\partial G^c_2}{\partial \tN_{ab}^d}
    \partial_c \tN_{ab}^d+
      \frac{\partial G^c_2}{\partial \tp^a} \partial_ 
      c\tp^a+
     \frac{\partial G^c_2}{\partial x^c}|_{exp}-\nonumber \\
&&      -\tN_{ab}^c\partial_c \tf^{ab}
-\tf^{ab}\partial_c \tN_{ab}^c-\tp^c\partial_c\tphi -
\tphi \partial_c \tp^c \:.\nonumber \\
\end{eqnarray}
Substituting the second type of generating function this into \rf{eqLagrangians} and then comparing the related terms yields a set of different equations
\begin{eqnarray}
\label{G2conditions}
&&\mH_J=\mH_E-\frac{\partial G_2^c}{\partial x^c}|_{expl} \ , \nonumber \\
&&\frac{\partial G_2^c}{\partial \tp^a}=\delta^c_a \tphi  \ , 
\quad \frac{\partial G_2^c}{\partial \tN^d_{ab}}=\tf^{ab}\delta_d^c  \ , \quad 
N^c_{ab}=\frac{\partial G_2^c}{f^{ab}} \ , \quad 
p^c=\frac{\partial G_2^c}{\partial B} \ . \nonumber \\
\end{eqnarray}
After this brief discussion of the canonical formalism, we proceed to the analysis of the question how Einstein and Jordan frame formulations of $F(R)$ gravity are related by canonical transformations. Recall that the
Covariant Hamiltonian in Jordan frame has the form \cite{Kluson:2020tzn}
\begin{align}
\label{HJ}
\mathcal{H}^J =& 
\frac{16\pi}{F'} f^{ab} \kzav{
N^c_{bd} N^d_{ac}
-\inv{3} N^c_{ac} N^d_{bd}
} \nonumber \\
&-\frac{8\pi}{3F'} f_{ab}
\kzav{\frac{F'}{F''}p^a - N^a_{cd} f^{cd}}
\kzav{\frac{F'}{F''}p^b - N^b_{ef} f^{ef}} \nonumber \\
&-\frac{\sqrt{-f}}{16\pi}\kzav{F-F'B}
\ . 
\end{align}
Using the transformations \rf{coordTransformation}, \rf{tNinJordan}, \rf{tpinJordan}, we can express the Einstein frame Hamiltonian in Jordan frame coordinates 
\begin{align}
\mathcal{H}^{E*} =& 
\frac{16\pi}{F'} f^{ab} \kzav{
N^c_{bd} N^d_{ac}
-\inv{3} N^c_{ac} N^d_{bd}
} \nonumber \\
&-\frac{8\pi}{3F'} f_{ab}
\kzav{\frac{F'}{F''}p^a - N^a_{cd} f^{cd}}
\kzav{\frac{F'}{F''}p^b - N^b_{ef} f^{ef}}\nonumber \\
&-\frac{\sqrt{-f}}{16\pi}\kzav{F-F'B}
\:,
\end{align}
which agrees with \rf{HJ}. The fact that these two Hamiltonians are equal means that  the   generating function of canonical transformation is independent on coordinates.
In fact, transformations
 \rf{coordTransformation}, \rf{tNinJordan} and \rf{tpinJordan} suggest that the generating function has the form of
\begin{equation}
\label{GeneratingFunctionG2}
G_2^c=F'\tN^c_{ab}f^{ab}+\sqrt{\frac{3}{2}}\ln F' \tp^c 
\:.
\end{equation}
In order to verify that the suggested form of $G_2^c$ is really the generating function of canonical transformations let us calculate the following derivative
\begin{eqnarray}
\frac{\partial G_2^c}{\partial B}=
F'' \tN_{ab}^c f^{ab} 
+\sqrt{\frac{3}{2}}\frac{F''}{F'}\tp^c
= \frac{F''}{F'} N_{ab}^c f^{ab} 
+\frac{F''}{F'}\kzav{\frac{F'}{F''} p^c - N_{ab}^c f^{ab} }
= p^c 
\end{eqnarray}
that agrees with the  the fourth equation in \rf{G2conditions}. Further, the derivative $G_2^c$ with respect to $f^{ab}$ gives
\begin{equation}
N^c_{ab}=\frac{\partial G_2^c}{\partial f^{ab}}=
F'\widetilde{N}^c_{ab}
\end{equation}
that gives the relation (\ref{tNinJordan}). Finally, the derivative of $G_2^c$ with respect to $\tp^a$ and $\widetilde{N}^d_{ab}$ lead to 
\begin{eqnarray}
&&\frac{\partial G_2^c}{\partial \tp^a}=\sqrt{\frac{3}{2}}\ln F'\delta^c_a=\delta^c_a\widetilde{\phi} \ , \nonumber \\
&& \frac{\partial G_2^c}{\partial \widetilde{N}^d_{ab}}=
F'\delta^c_d f^{ab}=\widetilde{f}^{ab}\delta^c_d
\nonumber \\
\end{eqnarray}
that lead to relations in (\ref{coordTransformation}).

In summary, we have shown that  \rf{GeneratingFunctionG2} is generating function of canonical transformation between Jordan and Einstein frame. This shows that these two frames are related by canonical transformations even in the case of covariant canonical formulations of these two theories which is new and non-trivial result.

\subsection{Remark About Poisson Brackets}\label{PoissonBrackets}
In Covariant Hamiltonian theory, the fundamental Poisson brackets are not always preserved under canonical transformation \cite{Struckmeier:2008zz}. Their purpose of canonical invariant is fulfilled with Lagrange brackets. This section summarizes Poisson brackets just for the reference. We will calculate Poisson brackets of Einstein frame variables in Jordan frame.
Since no Einstein frame coordinate depends on any of Jordan frame momenta \rf{coordTransformation}, the Poisson brackets of coordinates are zero. For the mixed brackets, one obtains
\begin{align}
\hzav{\widetilde{f}^{ab} \:,\: \widetilde{N}^c_{de}}_j
&=
\frac{\partial \widetilde{f}^{ab}}{\partial f^{ik}}\frac{\partial \widetilde{N}^c_{de}}{\partial N^j_{ik}}
+ \frac{\partial \widetilde{f}^{ab}}{\partial B}\frac{\partial \widetilde{N}^c_{de}}{\partial p^j}
- \frac{\partial \widetilde{N}^c_{de}}{\partial f^{ik}}\frac{\partial \widetilde{f}^{ab}}{\partial N^j_{ik}}
- \frac{\partial \widetilde{N}^c_{de}}{\partial B }\frac{\partial \widetilde{f}^{ab}}{\partial p^j}
= \delta^c_j \delta^{ab}_{cd}
\:,
\nonumber \\
\hzav{\widetilde{f}^{ab} \:,\: \widetilde{p}^c}_j
&=
\frac{\partial \widetilde{f}^{ab}}{\partial f^{ik}}\frac{\partial \widetilde{p}^c}{\partial N^j_{ik}}
+ \frac{\partial \widetilde{f}^{ab}}{\partial B}\frac{\partial \widetilde{p}^c}{\partial p^j}
- \frac{\partial \widetilde{p}^c}{\partial f^{ik}}\frac{\partial \widetilde{f}^{ab}}{\partial N^j_{ik}}
- \frac{\partial \widetilde{p}^c}{\partial B }\frac{\partial \widetilde{f}^{ab}}{\partial p^j}
= 0
\:,
\nonumber \\
\hzav{\widetilde{\phi} \:,\: \widetilde{N}^c_{de}}_j
&=
\frac{\partial \widetilde{\phi}}{\partial f^{ik}}\frac{\partial \widetilde{N}^c_{de}}{\partial N^j_{ik}}
+ \frac{\partial \widetilde{\phi}}{\partial B}\frac{\partial \widetilde{N}^c_{de}}{\partial p^j}
- \frac{\partial \widetilde{N}^c_{de}}{\partial f^{ik}}\frac{\partial \widetilde{\phi}}{\partial N^j_{ik}}
- \frac{\partial \widetilde{N}^c_{de}}{\partial B }\frac{\partial \widetilde{\phi}}{\partial p^j}
= 0
\:,
\nonumber \\
\hzav{\widetilde{\phi} \:,\: \widetilde{p}^c}_j
&=
\frac{\partial \widetilde{\phi}}{\partial f^{ik}}\frac{\partial \widetilde{p}^c}{\partial N^j_{ik}}
+ \frac{\partial \widetilde{\phi}}{\partial B}\frac{\partial \widetilde{p}^c}{\partial p^j}
- \frac{\partial \widetilde{p}^c}{\partial f^{ik}}\frac{\partial \widetilde{\phi}}{\partial N^j_{ik}}
- \frac{\partial \widetilde{p}^c}{\partial B }\frac{\partial \widetilde{\phi}}{\partial p^j}
= \delta^c_j
\:.
\nonumber \\
\end{align}
As we can see all mixed brackets yield the expected results. The brackets of momenta are those who break canonical invariance of Poisson brackets
\begin{align}
&\hzav{\widetilde{N}^f_{ab} \:,\: \widetilde{N}^c_{de}}_j
=
\frac{\partial \widetilde{N}^f_{ab}}{\partial f^{ik}}\frac{\partial \widetilde{N}^c_{de}}{\partial N^j_{ik}}
+ \frac{\partial \widetilde{N}^f_{ab}}{\partial B}\frac{\partial \widetilde{N}^c_{de}}{\partial p^j}
- \frac{\partial \widetilde{N}^c_{de}}{\partial f^{ik}}\frac{\partial \widetilde{N}^f_{ab}}{\partial N^j_{ik}}
- \frac{\partial \widetilde{N}^c_{de}}{\partial B }\frac{\partial \widetilde{N}^f_{ab}}{\partial p^j}
= 0
\:,
\nonumber \\
&\hzav{\widetilde{N}^d_{ab} \:,\: \widetilde{p}^c}_j
=
\frac{\partial \widetilde{N}^d_{ab}}{\partial f^{ik}}\frac{\partial \widetilde{p}^c}{\partial N^j_{ik}}
+ \frac{\partial \widetilde{N}^d_{ab}}{\partial B}\frac{\partial \widetilde{p}^c}{\partial p^j}
- \frac{\partial \widetilde{p}^c}{\partial f^{ik}}\frac{\partial \widetilde{N}^d_{ab}}{\partial N^j_{ik}}
- \frac{\partial \widetilde{p}^c}{\partial B }\frac{\partial \widetilde{N}^d_{ab}}{\partial p^j}
= \nonumber \\
&\qquad =
\sqrt{\frac{2}{3}}\inv{F'}\kzav{\delta^d_j N^c_{ab} - \delta^c_j N^d_{ab}}
\:,
\nonumber \\
&\hzav{\widetilde{p}^a \:,\: \widetilde{p}^c}_j
=
\frac{\partial \widetilde{p}^a}{\partial f^{ik}}\frac{\partial \widetilde{p}^c}{\partial N^j_{ik}}
+ \frac{\partial \widetilde{p}^a}{\partial B}\frac{\partial \widetilde{p}^c}{\partial p^j}
- \frac{\partial \widetilde{p}^c}{\partial f^{ik}}\frac{\partial \widetilde{p}^a}{\partial N^j_{ik}}
- \frac{\partial \widetilde{p}^c}{\partial B }\frac{\partial \widetilde{p}^a}{\partial p^j}
= \nonumber \\ 
&\qquad=
\frac{2}{3}\delta^c_j \kzav{
f^{ik} N^a_{ik} + \frac{F'(F''^2 - F'F''')}{F''^3} p^a
}
- \frac{2}{3}\delta^a_j \kzav{
f^{ik} N^c_{ik} + \frac{F'(F''^2 - F'F''')}{F''^3} p^c
}
\:. \nonumber \\
\end{align}
Contrary to the conventional Hamiltonian theory, the canonical transformation does not preserve fundamental Poisson brackets for momenta.

\section{Surface Lagrangian and Thermodynamic Properties}\label{fourth}
It is also important to mention the so-called surface part of the Lagrangian. This is the part which can be expressed as a derivative, or alternatively said as a divergence of some vector potential, and usually is neglected since it does not contribute to the equations of motion. However, this is not true just for any surface and more importantly, this part of Lagrangian contains information about thermodynamic properties of the boundary region, typically a horizon. Previously, we have found the surface Lagrangian as \rf{Lagrangians}
\begin{equation}
 \mathcal{L}_{sur} =  
\partial_i \hzav{\frac{\sqrt{-\widetilde{g}}}{16\pi}
\kzav{\widetilde{g}^{ab}\widetilde{\Gamma}^i_{ab} - \widetilde{g}^{ai}\widetilde{\Gamma}^b_{ab} 
+  \sqrt{6} \widetilde{\partial}^i \widetilde{\phi} }
} \ . 
\end{equation}
It is natural to express this surface term using canonical 
coordinates when we use the relations
\rf{inverseMomenta} and we get
\begin{equation}
\label{L_sur_Einstein}
  \mathcal{L}_{sur} =  
-\partial_i \kzav{
\widetilde{f}^{ab} \widetilde{N}^i_{ab} + \sqrt{\half{3}} \widetilde{p}^i
} \:.
\end{equation}
If we used coordinate transformation and put it back into the Jordan Frame, we would obtain
\begin{equation}
\label{L_sur_Jordan}
\mathcal{L}_{sur} = - \partial_i \kzav{ \frac{F'}{F''} p^i
}
\:,
\end{equation}
which is the same result, as was derived for Jordan Frame formulation. Thus, the thermodynamic properties can be assumed to be the same as in Jordan case \cite{Matous:2021uqh}, which is the expected result.

In General Relativity, there is a relation between the two Lagrangians (bulk and surface) \cite{Mukhopadhyay:2006vu}
\begin{equation}
\mL_{sur} = - \partial_c \kzav{
g_{ab} \pd{\mL_{bulk}}{\partial_c g_{ab}}
} \:,    
\end{equation}
which is better for our purposes to be written in $f-N$ formalism \cite{Parattu:2013gwa}
\begin{equation}\label{mL_sur}
\mL_{sur} = - \partial_c \kzav{ f^{ab} N^c_{ab} }
\:,    
\end{equation}
as visible from surface Lagrangian of Jordan \rf{L_sur_Jordan} as well as Einstein \rf{L_sur_Einstein} frame, this relation does not hold for $F(R)$ gravity theory. One could argue that \rf{mL_sur} is a kind of first approximation of more general formula, however its form is non-trivial and so it remains an open question.

{\bf Acknowledgement:}
\\
The work of J. Kluso\v{n}
is supported by the grant “Integrable Deformations”
(GA20-04800S) from the Czech Science Foundation
(GACR).

\appendix
\section{Equations of Motion}
This appendix summarizes the equations of motion. They are obtained from the Hamiltonian in a similar manner as it is done in the conventional Hamiltonian theory. The only difference rises from the unequal number of dimensions of coordinates and their conjugated momenta. The momenta always have dimension one higher. The equations of motion do not reveal all possible derivatives of momenta but only such which contracts the additional index \cite{Struckmeier:2008zz}. This leaves a bit of freedom for the momentum expression where two different momenta can express the same system. The presented calculation of equations of motion is very straight forward with the exception of equation for momentum $\tN^c_{ab}$, because its conjugated coordinate is included in the Hamiltonian \rf{EFHam} in three different forms --- $\tf^{ab}$, $\tf_{ab}$, $\tf$. To make that calculation easier, the following equations are handy
\begin{equation}
 \frac{\delta \widetilde{f}_{ik}}{\delta \widetilde{f}^{ab}} = 
 -\halb \widetilde{f}_{ak} \widetilde{f}_{bi}
  -\halb \widetilde{f}_{ai} \widetilde{f}_{bk}
  \:, \qquad
  \dd\widetilde{f} = \widetilde{f} \widetilde{f}_{ik} \dd \widetilde{f}^{ik}
\:.
\end{equation}
 Then the equations of motion for Einstein frame covariant Hamiltonian follow as
\begin{align}
&\partial_c \widetilde{f}^{ab} =
\frac{\delta \mathcal{H}^E}{\delta \widetilde{N}^c_{ab}}
 = 16\pi
\kzav{
\widetilde{f}^{ka} \widetilde{N}^b_{kc}
+ \widetilde{f}^{kb} \widetilde{N}^a_{kc}
- \inv{3} \widetilde{f}^{ka} \widetilde{N}^m_{km} \delta^b_c
- \inv{3} \widetilde{f}^{kb} \widetilde{N}^m_{km} \delta^a_c
}
\:, \nonumber\\
&\partial_c \widetilde{N}^c_{ab} =
-\frac{\delta \mathcal{H}^E}{\delta \widetilde{f}^{ab}}
 = 
 16\pi
\kzav{
\inv{3} \widetilde{N}^k_{ak} \widetilde{N}^m_{mb}
- \widetilde{N}^k_{am} \widetilde{N}^m_{kb}
}
-\inv{32 \pi}\sqrt{-\widetilde{f}}\widetilde{f}_{ab} V
-4\pi \widetilde{f}_{ai} \widetilde{f}_{bk} \widetilde{p}^i \widetilde{p}^k
\:, \nonumber\\
&\partial_a \widetilde{\phi}=
\frac{\delta \mathcal{H}^E}{\delta \widetilde{p}^{a}}
 = -8\pi \widetilde{f}_{ai} \widetilde{p}^i
\:, \nonumber\\
&\partial_c \widetilde{p}^c =
-\frac{\delta \mathcal{H}^E}{\delta \widetilde{\phi}}
 = -\inv{16\pi} \sqrt{-\widetilde{f}} V'
\:. \nonumber\\
\end{align}
\end{document}